\documentclass[letterpaper]{JHEP3}
\usepackage{
amsfonts,
epsfig,
amsmath,
xspace,
amssymb,
graphicx
}

\newcommand{\CH}{\mathcal H}
\DeclareMathOperator{\Tr}{Tr}

\preprint{\arXivid{0704.3719 [hep-th]} \\ SU-ITP-07/08 \\ KUNS-2069}

\title{A holographic proof of the strong subadditivity of entanglement entropy}

\author{Matthew Headrick \\ Stanford Institute for Theoretical Physics, Stanford,
CA 94305-4060, USA \\ \email{headrick@stanford.edu}}

\author{Tadashi Takayanagi\\ Department of Physics, Kyoto University, Kyoto, 606-8502,
Japan \\ \email{takayana@gauge.scphys.kyoto-u.ac.jp}}

\abstract{
When a quantum system is divided into subsystems, their entanglement entropies are subject to an inequality known as \emph{strong subadditivity}. For a field theory this inequality can be
 stated as follows: given any two regions of space $A$ and $B$, $S(A)+S(B)\ge S(A\cup B)+S(A\cap B)$.
 Recently, a method has been found for computing entanglement entropies
 in any field theory for which there is a holographically dual gravity theory.
 In this note we give a simple geometrical proof
of strong subadditivity employing this holographic prescription.}

\begin{document}

Entanglement entropy is an important tool in the study of quantum
information (see e.g.\ \cite{MR1796805}). It quantifies the extent to
which the state of a given subsystem of a quantum system
 is correlated with that of the rest of the system. Entanglement entropy enjoys
  a crucial mathematical property called \emph{strong subadditivity} \cite{MR0373508,MR0345558}.
   Recently, Ryu and one of the authors of the present paper
   \cite{Ryu:2006bv,Ryu:2006ef} proposed a relationship,
    applicable to any quantum field theory with a holographic gravity dual, between the
    entanglement entropy of a region of space in the QFT and the area of a certain minimal
    surface in the dual spacetime. An important test of the validity of this proposal is whether
     it has the property of strong subadditivity. This question was investigated in a variety
      of examples in the paper \cite{Hirata:2006jx}, always with an affirmative answer. In this
      note we give a general argument that it does, based only on general properties of holographic
       dualities. Besides giving support to the proposal, our argument provides
        an intuitive, geometrical way to understand strong subadditivity, a property
         whose formal algebraic proof is highly non-trivial.

The same argument can be used to establish a concavity property for
holographically-computed Wilson loop expectation values when the loops
are coplanar. This is briefly discussed in the last section of the paper.

\section{Strong subadditivity}

The von Neumann entropy of a density matrix $\rho$,
\begin{equation}
S(\rho) = -\Tr(\rho\ln\rho)\,,
\end{equation}
quantifies the extent to which the state represented by $\rho$ fails to be a pure state.
 If $\rho$ is obtained by tracing over part of the Hilbert space representing a subsystem---for example,
 one that is inaccessible to the experimentalist---then $S(\rho)$ is referred to as the
  \emph{entanglement entropy} of the remaining subsystem. More formally, if the Hilbert space
   of the full system factorizes into Hilbert spaces of two subsystems, $\CH_{\rm full}=\CH_1\otimes\CH_2$,
    then for each subsystem we define a reduced density matrix $\rho_1=\Tr_{\CH_2}\rho_{\rm full}$,
     $\rho_2=\Tr_{\CH_1}\rho_{\rm full}$, and a corresponding entanglement entropy $S(\rho_1)$
     and $S(\rho_2)$. On the basis of the concavity of the function $-x\ln x$ and elementary properties
      of Hilbert spaces, these can be shown quite generally to obey the following inequalities,
\begin{equation}\label{subadditivity}
|S(\rho_1)-S(\rho_2)| \le S(\rho_{\rm full}) \le S(\rho_1)+S(\rho_2)\,.
\end{equation}
This property of entanglement entropy is known as \emph{subadditivity} (the first inequality is also called the Araki-Lieb inequality \cite{Araki:1970ba}). In particular,
 if the full system is in a pure state then the two subsystems have the same entanglement entropy.

Now suppose the system is made up of more than two subsystems, $\CH_{\rm full} = \bigotimes_i\CH_i$.
 Then the inequalities \eqref{subadditivity} can be strengthened to yield \cite{MR0373508,MR0345558}
\begin{equation}\label{SSA}
S(\rho_{12})+S(\rho_{23}) \ge S(\rho_2) + S(\rho_{123})\,, \qquad
S(\rho_{12})+S(\rho_{23}) \ge S(\rho_1) + S(\rho_3)\,,
\end{equation}
where $\rho_{12}$ is the reduced density matrix for
$\CH_1\otimes\CH_2$, etc. These two inequalities can be shown to be
equivalent by the formal device of adding a fourth subsytem such
that $\rho_{1234}$ is a pure state. This property of entanglement
entropy is known as \emph{strong subadditivity}, and its proof is
highly non-trivial (although again it depends only on elementary
properties of Hilbert spaces).\footnote{Alternative proofs,
pedagogical expositions, and reviews can be found in
\cite{MR868631,MR1796805,MR1924445,Ruskai2004,MR2167530,2006quant.ph..4206R}.}
Strong subadditivity represents the concavity of the von Neumann
entropy and is a sufficiently strong property that it essentially
uniquely characterizes
 the von Neumann entropy \cite{MR0339901,MR0434290,MR0496300}.

In the context of a quantum field theory, a natural type of
subsystem to consider is that associated with a given region of
space. To any region $A$ is associated a Hilbert space $\CH_A$, and
for two disjoint regions $A$ and $B$ we have $\CH_{A\cup B}=\CH_A\otimes\CH_B$. For the entanglement entropy associated to $\CH_A$ we will write simply $S(A)$ (rather than $S(\rho_A)$). Due to the infinite number of degrees of freedom involved in a field theory, $S(A)$ typically suffers from an ultraviolet divergence proportional to the surface area of $A$ \cite{Bombelli:1986rw,Srednicki:1993im}. In order to deal with finite quantities one must impose a UV cutoff (and, if the surface area of $A$ is infinite, an IR cutoff as well). One may also consider subtracted quantities that remain finite as the UV cutoff is removed, such as the \emph{mutual information},
\begin{equation}
I(A,B) = S(A) + S(B) - S(A\cup B),
\end{equation}
defined when $A$ and $B$ (and their surfaces) are disjoint. By \eqref{subadditivity} this is non-negative. By employing these quantities, Casini and Huerta \cite{Casini:2004bw} showed that an analogue of the c-theorem in two dimensional QFTs can be derived from strong subadditivity.

To avoid confusion, it is important to remember that the concept of entanglement entropy
 refers to a specific state of the system at a specific time. Therefore all of the regions and
  surfaces we consider in this paper are restricted to a fixed constant-time slice of the field
   theory's spacetime.

\section{Holographic entanglement entropy}

\EPSFIGURE{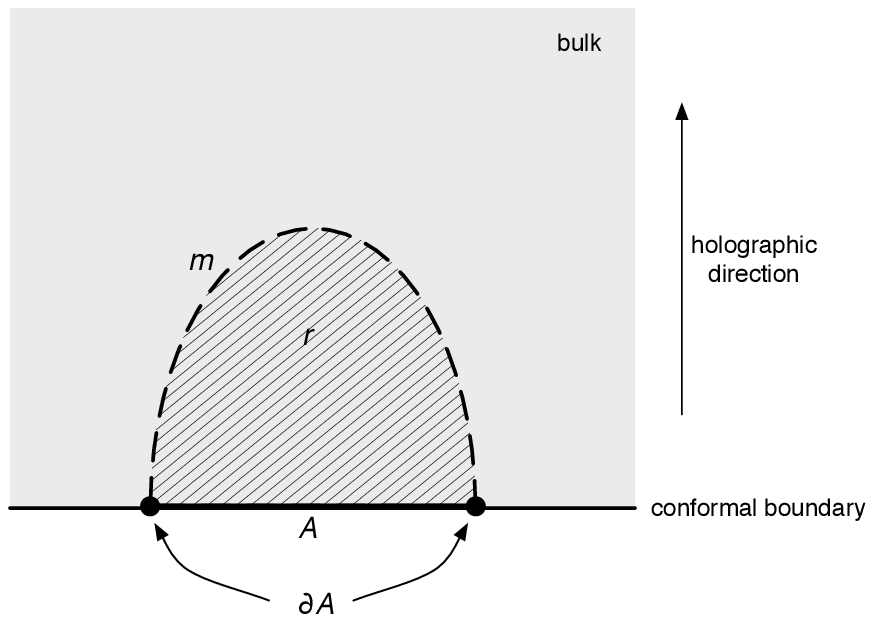,width=5in}{A constant-time slice of a spacetime on which
a gravity theory lives, and the conformal boundary on which its holographically dual field theory
 lives. $A$ is a region of the boundary; $m$ is the minimal hypersurface in the
 bulk ending on $\partial A$; and $r$ is a region of the bulk such that $\partial r = A\cup m$.\label{fig:SSA1}}

Recently, a proposal has been made in \cite{Ryu:2006bv,Ryu:2006ef}
for how to compute
 the entanglement entropy of a region of space in any quantum field theory that admits a holographic
  gravity dual. The proposal is very simple. The gravity theory lives in a space which as usual we call
   the bulk, and the QFT on its conformal boundary. (To avoid confusion, we will reserve the term
    ``boundary" for the space on which the QFT lives, and use the term ``surface" for the boundaries
     of various regions in the bulk and boundary.) We consider all
     hypersurfaces\footnote{We use the term ``hypersurface" because
     $m$ is \emph{spatially} co-dimension 1; in spacetime $m$ is of course co-dimension 2.} $m$ in the bulk that
      end on $\partial A$, and ask for the one with minimal area. (See Figure \ref{fig:SSA1}.) We then have
\begin{equation}\label{proposal}
S(A) = \frac1{4G_{\rm N}}\min_{m:\partial m=\partial A}a(m)\,,
\end{equation}
where $a(m)$ is the area of $m$. For the case when the bulk gravity theory lives
on a static asymptotically AdS spacetime, Fursaev
\cite{Fursaev:2006ih} has given a derivation of
 \eqref{proposal} using Euclidean Quantum Gravity and the basic
 principles \cite{Gubser:1998bc, Witten:1998qj} of the AdS/CFT correspondence
 \cite{Maldacena:1997re}. Notice that the expression (\ref{proposal}) coincides
  with the Bekenstein-Hawking formula of black hole entropy if we replace the
   minimal surface with a black hole horizon. Indeed, at high temperature the
    spacetime of the gravity theory generally includes a horizon; when part of
     the minimal surface wraps the horizon, its contribution corresponds to the
      usual thermal entropy. We will see an example of this situation when we come
       to Figure \ref{fig:SSA2} below.

Three refinements should be made to \eqref{proposal}. First, both sides are divergent; the
 left-hand side is ultraviolet divergent as discussed above, while the right-hand side is
  infrared divergent due to the infinite proper distance from any point in the bulk to the
   conformal boundary. It is easy to see that the latter divergence, like the former, is proportional
    to the surface area of $A$. In fact, these two divergences are the same, a manifestation of the usual
     UV/IR correspondence characteristic of holographic dualities. Therefore \eqref{proposal} is
      meant to apply in the presence of a UV cutoff in the QFT, corresponding to an IR cutoff in
       the gravity theory. The simplest such cutoff is a brute force one that cuts off the bulk space
        at a finite value of the holographic coordinate. The exact choice of cutoff will not be
         important in what we say below, and for simplicity of presentation we will leave it implicit
         in the discussion.

\EPSFIGURE{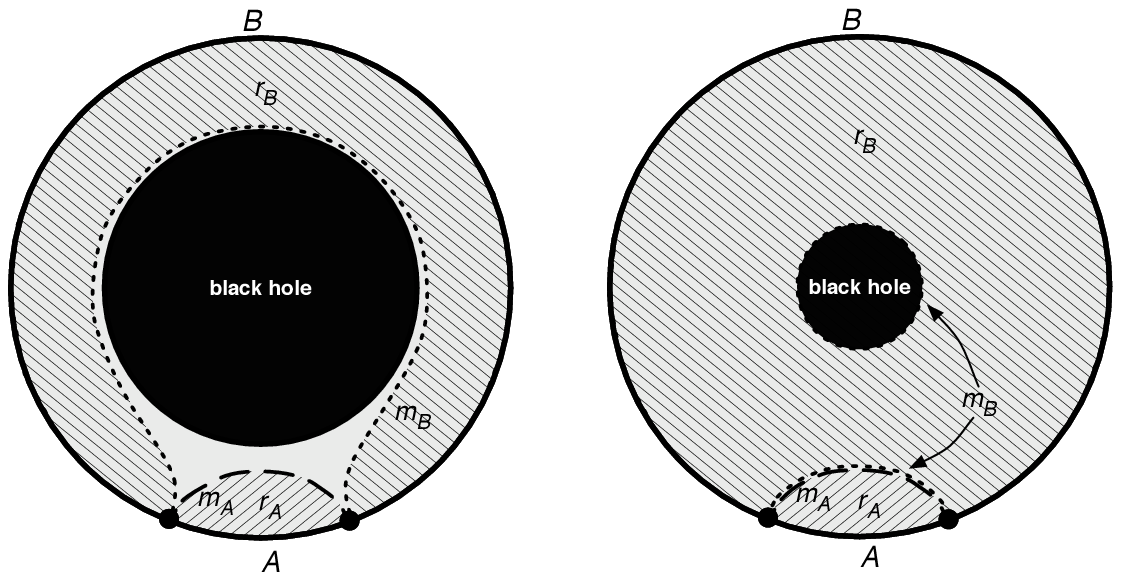,width=6.1in}{
\label{fig:SSA2}
Two examples in which the QFT lives on a compact space and the bulk contains a black hole.
 (Technically, since we are considering an eternal black hole in static coordinates, in each
 case the full spacetime consists of two copies of the region shown connected by an Einstein-Rosen
  bridge; this won't affect our discussion.) The boundary is divided into two regions $A$ and $B$.
   Since $\partial A=\partial B$, if the bulk had trivial $(d-1)$st homology the corresponding minimal
    hypersurfaces $m_A$ and $m_B$ would be identical, and we would have $S(A)=S(B)$. However, due
     to the requirement that each hypersurface be homologous to the corresponding boundary
      region, $m_B$ can either (left) wrap around the other side of the event horizon, or (right)
       separate into two connected components, one being $m_A$ and the other the event horizon.
        In the latter case we have $S(B)=S(A)+S_{\rm BH}$, where $S_{\rm BH}$ is the black hole's
         Bekenstein-Hawking entropy; since $S_{\rm full} = S_{\rm BH}$, the Araki-Lieb inequality is saturated.
}

Second, there is a complication that occurs when the bulk has non-trivial $(d-1)$st homology
(where $d$ is the spatial dimension of the bulk, which is also the spacetime dimension of the
 boundary). This will be the case, for example, when the bulk contains a black hole. Fursaev's
  derivation of \eqref{proposal} then tells us that we should minimize $a$ not over all hypersurfaces
   ending on $\partial A$ but only over those that are homologous to $A$; that is, there should exist
    a region $r$ of the bulk such that $\partial r=A\cup m$. See Figures \ref{fig:SSA1} and \ref{fig:SSA2}
     for examples. (See \cite{Ryu:2006bv,Ryu:2006ef,Emparan:2006ni,Fursaev:2006ih} for further discussion.)
      This rule will be essential in what follows.

Third, the formula \eqref{proposal} is exact in the limit that the gravity in the bulk
 is controlled by the Einstein-Hilbert action. Higher curvature corrections to the bulk
  action will lead to corrections to the functional $a(m)$. For example, Fursaev \cite{Fursaev:2006ih}
   showed that, if the bulk action is corrected by a Gauss-Bonnet term, then $a(m)$ is corrected by
    an Einstein-Hilbert term,
\begin{equation}\label{withEH}
a(m) = \int_m\sqrt{h}\left(1 + 2\alpha R(h)\right) + 4\alpha\int_{\partial m}\sqrt{\gamma}K\,,
\end{equation}
where $h$ is the induced metric on $m$ and $\alpha$ is the coefficient of the Gauss-Bonnet term in
the bulk action (see \cite{Fursaev:2006ih} for details). In order to make the variational
 problem for $m$ well defined, we have also included a Gibbons-Hawking boundary term; $\gamma$
  is the induced metric on $\partial m$, and $K$ is the trace of its extrinsic curvature (in $m$).

All of these regions and surfaces---both on the boundary and in the bulk---must lie on a single
constant-time slice. In order to have a well-defined notion of ``constant-time slice" in the bulk,
 we must restrict ourselves to states for which the bulk geometry is static.
 A covariant
 generalization of \eqref{proposal} to time-dependent geometries
 will be discussed in \cite{timedependent}. We leave the proof of
 strong subadditivity in that context to future work.

\section{Proof}

In the paper \cite{Hirata:2006jx} the authors investigated in a
variety of examples
 whether the formula \eqref{proposal} for the entanglement entropy satisfied the property of
  strong subadditivity, and in all cases studied it did. Here we will give a
   simple argument that
   it does in general.

We begin by rewriting the inequalities \eqref{SSA} in the forms
\begin{equation}\label{ABform}
S(A) + S(B) \ge S(A\cup B)+S(A\cap B)\,, \qquad S(A) + S(B) \ge
S(A\setminus B)+S(B\setminus A)\,,
\end{equation}
where $A\setminus B\equiv A\cap B^{\rm c}$. We will prove the first inequality; the proof of the
 second one is very similar and is left as an exercise to the reader.

\EPSFIGURE{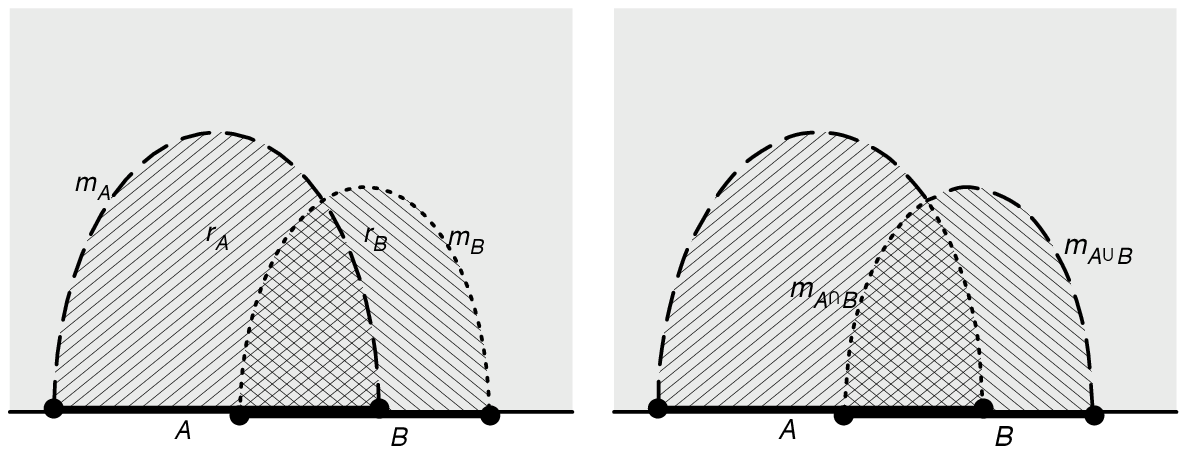,width=6.1in}{
\label{fig:SSA3}
Two overlapping regions $A$ and $B$ of the boundary, with (left) their respective minimal bulk
 hypersurfaces $m_A$, $m_B$ and bulk regions $r_A$, $r_B$, and (right) their minimal hypersurfaces
  $m_A$ and $m_B$ cut up and rearranged into two new hypersurfaces $m_{A\cup B}$ (the bulk part of
   the surface of $r_A\cup r_B$) and $m_{A\cap B}$ (the bulk part of the surface of $r_A\cap r_B$).
    $m_{A\cup B}$ and $m_{A\cap B}$ end on $\partial(A\cup B)$ and $\partial(A\cap B)$ respectively
    (although they are not necessarily the minimal such hypersurfaces).
}

Let $m_A$, $m_B$ be the minimal hypersurfaces
 in the bulk ending on $\partial A$, $\partial B$ respectively, and $r_A$, $r_B$ be the
 corresponding regions of the bulk (so that $\partial r_A = A\cup m_A$, $\partial r_B = B\cup m_B$).
  (See Figure \ref{fig:SSA3}, left side.) We now define the regions
\begin{equation}
r_{A\cup B} = r_A\cup r_B\,, \qquad r_{A\cap B} = r_A\cap r_B\,.
\end{equation}
We can decompose the surfaces of these regions as usual into a part on the boundary and a part in the bulk,
\begin{equation}
\partial r_{A\cup B} = (A\cup B)\cup m_{A\cup B}, \qquad \partial r_{A\cap B} = (A\cap B)\cup m_{A\cap B}\,.
\end{equation}
(See Figure \ref{fig:SSA3}, right side.) Clearly $m_{A\cup B}$ ends on $\partial(A\cup B)$. While nothing says that
 it is the \emph{minimal} hypersurface ending on $\partial(A\cup B)$, its area is an upper bound on the area of
  the minimal one, and therefore on $4G_{\rm N}\,S(A\cup B)$; similarly for $A\cap B$. Now the
   hypersurfaces $m_{A\cup B}$ and $m_{A\cap B}$ are simply rearrangements of $m_A$ and $m_B$
    (meaning that $m_{A\cup B}\cup m_{A\cap B} = m_A\cup m_B$), so they have the same total
     areas,\footnote{In this sentence we've assumed the generic situation that $m_A$ and $m_B$
      intersect along (spatially) codimension 2 submanifolds. More generally we have
       $m_{A\cup B}\cup m_{A\cap B} \subset m_A\cup m_B$ and $a(m_{A\cup B}) + a (m_{A\cap B})
        \le a(m_A) + a(m_B)$.}
\begin{equation}\label{rearrangement}
a(m_{A\cup B}) + a (m_{A\cap B}) = a(m_A) + a(m_B)\,,
\end{equation}
which completes the proof.

Note that equation \eqref{rearrangement} holds not just if $a$ is the area, but if it is any
\emph{extensive} functional of the hypersurface. This means that if $m$ and $m'$ are
 two disjoint hypersurfaces with a common boundary, $\partial m \cap \partial m'\neq\emptyset$, then we have
  $a(m\cup m') = a(m) + a(m')$. This is true, for example, for the
  Einstein-Hilbert term (with boundary term) added in \eqref{withEH}.

\section{Discussion}

In this letter we gave a simple geometric proof of strong
subadditivity of entanglement entropy based on the holographic
formula (\ref{proposal}). The extra dimension in the holographic
dual obviously plays an essential role in this proof. Since the
strong subadditivity of entanglement entropy should be true in any
quantum mechanical many-body system, our result shows that the idea
of holography is consistent with any quantum system from this basic
viewpoint.

It is interesting to ask when the inequalities \eqref{SSA} are saturated. The only examples we know in the holographic context involve only two disjoint regions, and therefore reduce to the saturation of weak subadditivity, inequalities \eqref{subadditivity}. (It would be interesting to find examples where this is not the case.) The first of these, the Araki-Lieb inequality, is obviously saturated when the full system is pure; then each entanglement entropy is due only to correlations between the subsystems, rather than to the full system being in a mixed state. A system that is in a mixed state but nonetheless saturates the Araki-Lieb inequality is depicted on the right side of Figure \ref{fig:SSA2}. The fact that $m_B$ is disconnected suggests that here the entanglement entropy of $B$ has two separate and unrelated origins: the thermal entropy of the full system ($S_{\rm full}$), and the correlations between $A$ and $B$ ($S(A)$). On the left side of that figure, where $m_B$ is connected, the inequality is not saturated.
 
As for the second inequality in \eqref{subadditivity}, it is saturated (i.e.\ the mutual information vanishes) when two regions are sufficiently far apart that their union's minimal hypersurface does not connect them but instead is simply the union of their respective minimal hypersurfaces. This was seen in explicit examples in \cite{Hirata:2006jx}. The mutual information vanishes if and only if the two systems are uncorrelated, i.e.\ $\rho_{12} = \rho_1\otimes\rho_2$ \cite{MR1796805}. It is interesting that the correlations can go strictly to zero in a field theory (in the large $N$ limit).

Finally, it is useful to notice that our argument can be directly
applied to the holographic derivation of a concavity property of
coplanar Wilson loops \cite{Hirata:2006jx}, which is closely related
to the Bachas inequality \cite{Bachas:1985xs}. If the curves
$C_A=\partial A$ and $C_B=\partial B$ lie in the same
two-dimensional plane, then it is clear that the holographically
computed expectation values of the corresponding Wilson loops
satisfy
\begin{eqnarray}\label{wilsonssa}
&&\langle W(C_A)\rangle \langle W(C_B)\rangle  \leq \langle
W(C_{A\cap B})\rangle \langle W(C_{A\cup B})\rangle\,, \\ && \langle
W(C_A)\rangle \langle W(C_B)\rangle  \leq \langle W(C_{A\setminus
B})\rangle \langle W(C_{B\setminus A})\rangle ,
\end{eqnarray}
where we defined $C_{A\cap B}=\partial(A\cap B)$ etc.
 They are equivalent to (\ref{ABform}) once we remember
that the holographic Wilson loop expectations can also be found from
the minimal surface \cite{Maldacena:1998im,Rey:1998ik}. The evidence from the
gauge theory side
for these relations will be discussed in \cite{hirata}.

\acknowledgments

M.H. is supported  by the Stanford Institute for Theoretical Physics
and by NSF grant PHY 9870115. T.T. would like to thank Stanford
Institute for Theoretical Physics for hospitality, where the present
work was initiated. T.T. is also grateful to T. Hirata, V. Hubeny,
M. Rangamani for useful comments on the draft of this paper. The
work of T.T. is supported in part by JSPS Grant-in-Aid for
Scientific Research No.18840027.

%
\bibliography{ref}

\providecommand{\href}[2]{#2}\begingroup\raggedright\begin{thebibliography}{10}

\bibitem{MR1796805}
M.~A. Nielsen and I.~L. Chuang, {\em Quantum computation and quantum
  information}.
\newblock Cambridge University Press, Cambridge, 2000.

\bibitem{MR0373508}
E.~H. Lieb and M.~B. Ruskai, {\it A fundamental property of quantum-mechanical
  entropy},  {\em Phys. Rev. Lett.} {\bf 30} (1973) 434--436.

\bibitem{MR0345558}
E.~H. Lieb and M.~B. Ruskai, {\it Proof of the strong subadditivity of
  quantum-mechanical entropy},  {\em J. Math. Phys.} {\bf 14} (1973)
  1938--1941. With an appendix by B. Simon.

\bibitem{Ryu:2006bv}
S.~Ryu and T.~Takayanagi, {\it Holographic derivation of entanglement entropy
  from {AdS/CFT}},  {\em Phys. Rev. Lett.} {\bf 96} (2006) 181602,
  [\href{http://xxx.lanl.gov/abs/hep-th/0603001}{{\tt hep-th/0603001}}].

\bibitem{Ryu:2006ef}
S.~Ryu and T.~Takayanagi, {\it Aspects of holographic entanglement entropy},
  {\em JHEP} {\bf 08} (2006) 045,
  [\href{http://xxx.lanl.gov/abs/hep-th/0605073}{{\tt hep-th/0605073}}].

\bibitem{Hirata:2006jx}
T.~Hirata and T.~Takayanagi, {\it {AdS/CFT} and strong subadditivity of
  entanglement entropy},  {\em JHEP} {\bf 02} (2007) 042,
  [\href{http://xxx.lanl.gov/abs/hep-th/0608213}{{\tt hep-th/0608213}}].

\bibitem{Araki:1970ba}
H.~Araki and E.~H. Lieb, {\it Entropy inequalities},  {\em Commun. Math. Phys.}
  {\bf 18} (1970) 160--170.

\bibitem{MR868631}
D.~Petz, {\it Quasi-entropies for finite quantum systems},  {\em Rep. Math.
  Phys.} {\bf 23} (1986), no.~1 57--65.

\bibitem{MR1924445}
M.~B. Ruskai, {\it Inequalities for quantum entropy: a review with conditions
  for equality}, .

\bibitem{Ruskai2004}
M.~B. Ruskai, {\it {Lieb's simple proof of concavity of $\Tr A^p K^* B^(1-p) K$
  and remarks on related inequalities}},
  \href{http://xxx.lanl.gov/abs/quant-ph/0404126v4}{{\tt quant-ph/0404126v4}}.

\bibitem{MR2167530}
M.~A. Nielsen and D.~Petz, {\it A simple proof of the strong subadditivity
  inequality},  {\em Quantum Inf. Comput.} {\bf 5} (2005), no.~6 507--513,
  [\href{http://xxx.lanl.gov/abs/quant-ph/0408130}{{\tt quant-ph/0408130}}].

\bibitem{2006quant.ph..4206R}
M.~B. {Ruskai}, {\it Another short and elementary proof of strong subadditivity
  of quantum entropy},  \href{http://xxx.lanl.gov/abs/quant-ph/0604206}{{\tt
  quant-ph/0604206}}.

\bibitem{MR0339901}
J.~Acz{\'e}l, B.~Forte, and C.~T. Ng, {\it Why the {S}hannon and {H}artley
  entropies are `natural'},  {\em Advances in Appl. Probability} {\bf 6} (1974)
  131--146.

\bibitem{MR0434290}
W.~Ochs, {\it A new axiomatic characterization of the von {N}eumann entropy},
  {\em Rep. Math. Phys.} {\bf 8} (1975), no.~1 109--120.

\bibitem{MR0496300}
A.~Wehrl, {\it General properties of entropy},  {\em Rev. Modern Phys.} {\bf
  50} (1978), no.~2 221--260.

\bibitem{Bombelli:1986rw}
L.~Bombelli, R.~K. Koul, J.-H. Lee, and R.~D. Sorkin, {\it A quantum source of
  entropy for black holes},  {\em Phys. Rev.} {\bf D34} (1986) 373.

\bibitem{Srednicki:1993im}
M.~Srednicki, {\it Entropy and area},  {\em Phys. Rev. Lett.} {\bf 71} (1993)
  666--669, [\href{http://xxx.lanl.gov/abs/hep-th/9303048}{{\tt
  hep-th/9303048}}].

\bibitem{Casini:2004bw}
H.~Casini and M.~Huerta, {\it A finite entanglement entropy and the c-theorem},
   {\em Phys. Lett.} {\bf B600} (2004) 142--150,
  [\href{http://xxx.lanl.gov/abs/hep-th/0405111}{{\tt hep-th/0405111}}].

\bibitem{Fursaev:2006ih}
D.~V. Fursaev, {\it Proof of the holographic formula for entanglement entropy},
   {\em JHEP} {\bf 09} (2006) 018,
  [\href{http://xxx.lanl.gov/abs/hep-th/0606184}{{\tt hep-th/0606184}}].

\bibitem{Gubser:1998bc}
S.~S. Gubser, I.~R. Klebanov, and A.~M. Polyakov, {\it Gauge theory correlators
  from non-critical string theory},  {\em Phys. Lett.} {\bf B428} (1998)
  105--114, [\href{http://xxx.lanl.gov/abs/hep-th/9802109}{{\tt
  hep-th/9802109}}].

\bibitem{Witten:1998qj}
E.~Witten, {\it Anti-de {S}itter space and holography},  {\em Adv. Theor. Math.
  Phys.} {\bf 2} (1998) 253--291,
  [\href{http://xxx.lanl.gov/abs/hep-th/9802150}{{\tt hep-th/9802150}}].

\bibitem{Maldacena:1997re}
J.~M. Maldacena, {\it The large {N} limit of superconformal field theories and
  supergravity},  {\em Adv. Theor. Math. Phys.} {\bf 2} (1998) 231--252,
  [\href{http://xxx.lanl.gov/abs/hep-th/9711200}{{\tt hep-th/9711200}}].

\bibitem{Emparan:2006ni}
R.~Emparan, {\it Black hole entropy as entanglement entropy: {A} holographic
  derivation},  \href{http://xxx.lanl.gov/abs/hep-th/0603081}{{\tt
  hep-th/0603081}}.

\bibitem{timedependent}
V.~Hubeny, M.~Rangamani, and T.~Takayanagi, ``A covariant holographic
  entanglement entropy proposal.''
\newblock To appear.

\bibitem{Bachas:1985xs}
C.~Bachas, {\it Convexity of the quarkonium potential},  {\em Phys. Rev.} {\bf
  D33} (1986) 2723.

\bibitem{Maldacena:1998im}
J.~M. Maldacena, {\it Wilson loops in large {N} field theories},  {\em Phys.
  Rev. Lett.} {\bf 80} (1998) 4859--4862,
  [\href{http://xxx.lanl.gov/abs/hep-th/9803002}{{\tt hep-th/9803002}}].

\bibitem{Rey:1998ik}
S.-J. Rey and J.-T. Yee, {\it Macroscopic strings as heavy quarks in large {N}
  gauge theory and anti-de {S}itter supergravity},  {\em Eur. Phys. J.} {\bf
  C22} (2001) 379--394, [\href{http://xxx.lanl.gov/abs/hep-th/9803001}{{\tt
  hep-th/9803001}}].

\bibitem{hirata}
T.~Hirata.
\newblock To appear.

\end{thebibliography}\endgroup
\bibliographystyle{JHEP}
%

\end{document}